\documentclass[journal]{IEEEtran}
\usepackage{amsmath,amsfonts}
\usepackage{algorithm}
\usepackage{algpseudocode}
\usepackage{array}
\usepackage[caption=false,font=normalsize,labelfont=sf,textfont=sf]{subfig}
\usepackage{textcomp}
\usepackage{stfloats}
\usepackage{url}
\usepackage{verbatim}
\usepackage{graphicx, booktabs}
\usepackage{cite}
\usepackage{color}
\usepackage{pifont}

\newcommand{\cmark}{\textcolor{green!60!black}{\scalebox{1.2}{\ding{51}}}} 
\newcommand{\xmark}{\textcolor{red}{\scalebox{1.2}{\ding{55}}}}            

\usepackage{xcolor}
\usepackage{multirow}
\usepackage{mathabx} 
\allowdisplaybreaks 

\begin{document}

\title{A TSO-DSO Coordination Framework via Analytical Representation and Monetization of PQV-Based Distribution System Flexibility}

\author{Burak~Dindar,~\IEEEmembership{Graduate Student Member,~IEEE,}
          ~Can~Berk~Saner,~\IEEEmembership{Member,~IEEE,}
          ~Hüseyin~K.~Çakmak,
          and~Veit~Hagenmeyer,~\IEEEmembership{Member,~IEEE}

        \vspace{-0.6 cm}
          
          \thanks{

          Burak Dindar, Hüseyin K. Çakmak and Veit Hagenmeyer are with the Institute for Automation and Applied Informatics, Karlsruhe Institute of Technology, 76131 Karlsruhe, Germany, (e-mail:burak.dindar@kit.edu; hueseyin.cakmak@kit.edu; veit.hagenmeyer@kit.edu).
          
          Can Berk Saner is with the Department of Mathematics, National University of Singapore, Singapore 119076, (e-mail: sanerc@u.nus.edu).  
}
}

\maketitle

\begin{abstract}
As the role of distribution system (DS) flexibility in transmission system operator (TSO) network management becomes increasingly vital, data privacy concerns hinder seamless interoperability. The notion of the feasible operating region (FOR), defined in the PQ domain, has emerged as a promising privacy-preserving approach. However, effectively leveraging FOR in TSO operations remains challenging due to three key factors: its accurate determination in large-scale, meshed DS networks; its tractable analytical representation; and its economic valuation. In the present paper, we propose a novel AC optimal power flow (OPF)-based method to construct a three-dimensional PQV-FOR, explicitly accounting for voltage variability and diverse flexibility-providing unit (FPU) characteristics. The construction process employs a two-stage sampling strategy that combines bounding box projection and Fibonacci direction techniques to efficiently capture the FOR. We then introduce an implicit polynomial fitting approach to analytically represent the FOR. Furthermore, we derive a quadratic cost function over the PQV domain to monetize the FOR. Thus, the proposed framework enables single-round TSO–DSO coordination: the DSO provides an analytical FOR and cost model; the TSO determines operating point at the point of common coupling (PCC) within the FOR-based AC-OPF; and the DSO computes FPU dispatch by solving its local OPF, without computationally intensive disaggregation or iterative coordination. Case studies on meshed DS with up to 533 buses, integrated into TS, demonstrates the method’s efficiency compared to standard AC-OPF. On average, the proposed approach yields negligible cost deviations of at most 0.058\% across test cases, while reducing computation times by up to 58.11\%.
\end{abstract}


\vspace{-0.6 cm}

\section{Introduction}

The power system is rapidly transforming under carbon neutrality targets, leading to a significant increase in the number of distributed generators (DGs). While DGs introduce considerable uncertainty, they also offer valuable distribution-level flexibility that can be leveraged for network management. These flexibility-providing units (FPUs) are predominantly connected to distribution systems (DSs), making the role of DSs increasingly critical in ensuring reliable and efficient grid operation. Consequently, enhanced coordination between Transmission System Operators (TSOs) and Distribution System Operators (DSOs) is essential for managing the overall power system effectively \cite{givisiez2020review}. However, this coordination is often hindered by stakeholders' concerns regarding data privacy \cite{mohammadi2018diagonal}. Therefore, privacy-preserving methodologies are crucial to enable secure and effective TSO-DSO interactions.

In this context, the Feasible Operating Region (FOR) approach, based on the PQ chart, has been proposed to represent the aggregated flexibility of the DS. The FOR captures all feasible combinations of active and reactive power at the point of common coupling (PCC), while respecting DS constraints such as voltage limits, thermal limits, and FPU operating ranges. By sharing this compact PQ representation instead of sensitive data (e.g., topology or customer information), TSOs and DSOs can coordinate effectively \cite{capitanescu2018tso}. Once the FOR is defined, it can be directly utilized by the TSO.

\vspace{-0.4 cm}

\subsection{Related Work}\label{sec:related_work}

In the literature, the FOR is computed using various techniques, including geometric methods (e.g., Minkowski sum), random sampling (RS), and optimization-based approaches, particularly those relying on optimal power flow (OPF). In geometric methods, the flexibilities of FPUs are geometrically aggregated to form the FOR \cite{kundu2018approximating, ozturk2022aggregation}. However, a major limitation is that it does not account for the underlying grid constraints \cite{contreras2021computing}. Alternatively, the RS-based approach involves generating a large number of random operating points, followed by power flow analyses to determine whether each point satisfies grid constraints. The collection of feasible operating points is then used to construct the FOR \cite{alsharif2023comparison}. While this method accounts for grid limitations, it is computationally intensive, as a substantial number of samples are required \cite{heleno2015estimation}. Due to these limitations, optimization-based methods have gained prominence as a more efficient solution \cite{sarstedt2021survey}.

Among OPF-based methods, one commonly used approach involves linearizing the power system model \cite{kalantar2019characterizing, fortenbacher2020reduced}. While this method offers computational efficiency, the linearization of nonlinear power flow equations leads to inaccuracies in the determination of the FOR \cite{sarstedt2021survey}. Another category is sampling strategies \cite{vijay2022feasibility}, which are typically classified into angle-based and set-point-based approaches. In these methods, the OPF is solved for varying objective function coefficients in the PQ domain \cite{silva2018estimating, lopez2021quickflex}. Compared to other techniques, these strategies can more effectively identify the boundary of the FOR \cite{contreras2021computing}. However, most existing studies apply these methods to radial test systems with a limited number of buses and often assume a fixed voltage at the PCC \cite{capitanescu2018tso, rabiee2024exploiting}. In practice, the operation of DSs as meshed networks is becoming increasingly prevalent, necessitating that these characteristics be accounted for in analysis. Moreover, to fully exploit the flexibility potential of DSs, the PCC voltage should not be assumed constant; instead, analyses should be conducted in the three-dimensional PQV domain \cite{stark2024novel}. Given these requirements, there is a need for innovative approaches to accurately determine the FOR.

After the FOR is generated, a critical question arises: \textit{how can the FORs computed by DSOs be effectively utilized by TSOs in their operational planning?} One common solution involves iterative coordination methods that establish a hierarchical interaction between TSOs and DSOs. In these frameworks, the DSO first determines the FOR, and the TSO independently solves its own OPF problem, subsequently proposing a power exchange operating point at the PCC. The DSO then verifies whether the proposed point lies within its FOR \cite{yoon2024cooperative, stark2023determination, liu2025region}. However, this process often requires a significant number of iterations, leading to high communication overhead. To mitigate this issue, it is desirable for the FOR to be represented analytically, enabling its direct integration into the TSO’s OPF formulation. Yet, due to the potentially non-convex nature of the FOR, achieving an accurate and tractable analytical representation remains challenging. As a result, several simplifying assumptions are typically employed in the literature. 

For example, \cite{contreras2021congestion} approximates the FOR as a polygon and integrates it into a linear OPF through a set of linear inequalities. Similarly, in \cite{capitanescu2018opf} and \cite{kalantar2024grid}, the FOR is modeled using a simple rectangular PQ box, which facilitates its incorporation but neglects the underlying complexity of the actual FOR. \cite{bandeira2024adp} employs the LinDistFlow model to compute a conservative approximation of the FOR, ensuring that the resulting bounds remain within the actual FOR, thus avoiding the risk of selecting infeasible setpoints—a common issue in convex hull-based approaches. Despite their practical utility, they often lead to over-conservative or inaccurate representations. Therefore, there is a pressing need for novel analytical representations that strike a balance between computational tractability and accurate depiction of the FOR's complex geometry.

Another important consideration is that the FOR solely characterizes the feasibility of operating points. However, for the FOR to be effectively integrated into the TSO's OPF problem, it is also necessary to monetize the aggregated flexibility within the PQV (or PQ) domain. This requires a representation of the cost of flexibility in an analytical form, analogous to the analytical FOR representation. Achieving this is particularly challenging due to the nonlinear nature of power system. One of the earliest efforts in this direction is presented in \cite{sarstedt2022monetarization}, where monetization is approached through an aggregation and disaggregation process. However, this method relies on a brute-force evaluation of the cost at each point on the FOR, resulting in significant computational burden. Similarly, \cite{churkin2023impacts} uses a brute-force method to assign costs across the FOR but does not provide an analytical cost representation.

A notable advancement is found in \cite{capitanescu2023computing}, where the cost of aggregated flexibility is represented analytically. In this study, piecewise linear cost functions are independently derived for active and reactive power. Despite this innovation, the approach is only tested on the 34-bus radial test system and is not validated through integration into an OPF formulation. In \cite{bozionek2022design}, the cost is approximated using a continuous quadratic function, but only for the reactive power, with the active power and voltage assumed constant. This significantly simplifies the problem but limits applicability to broader flexibility modeling. Meanwhile, \cite{bandeira2024adp} emphasizes that the variables at the PCC, including P, Q, and V, must be considered holistically when defining the cost function. This study fits the cost using a quadratic function, but relies on the simplified LinDistFlow model and is validated only on a small 15-bus radial test system. These studies collectively highlight that, to fully leverage DS flexibility within TSO operations, the cost of the FOR must be represented in a form that is both analytically tractable and compatible with OPF formulations. Furthermore, to ensure practical applicability, these approaches must be extended to more realistic conditions, including meshed network topologies, large-scale systems, and AC-OPF. An overview of the related work is given in Table \ref{tab:literature}.

\begin{table}[]
\centering
\caption{Overview of the Related Work.}
\label{tab:literature}
\resizebox{\columnwidth}{!}{%
\begin{tabular}{@{}ccccccc@{}}
\toprule
Work &
\begin{tabular}[c]{@{}c@{}}Uses Full\\ AC-PF Model \end{tabular} &
\begin{tabular}[c]{@{}c@{}}Meshed DS\end{tabular} &
\begin{tabular}[c]{@{}c@{}}Considered\\ Variables at PCC\end{tabular} &
\begin{tabular}[c]{@{}c@{}} Analytical FOR\end{tabular} &
\begin{tabular}[c]{@{}c@{}}Analytical\\ Cost\end{tabular} &
\begin{tabular}[c]{@{}c@{}}TSO-DSO \\ Coord. Scheme\end{tabular} \\ \midrule
\cite{yoon2024cooperative} & \xmark & \xmark & P, Q & \cmark (Polygon) & \xmark & Iterative  \\
\cite{stark2023determination} & \cmark & \cmark & P, Q & \xmark & \xmark & Iterative \\
\cite{liu2025region} & \xmark & \xmark & P, Q & \cmark (Polygon) & \xmark & Iterative \\
\cite{contreras2021congestion} & \xmark & \xmark & P, Q & \cmark (Polygon) & \xmark & Iterative \\
\cite{capitanescu2018opf} & \cmark & NS & P, Q & \cmark (PQ Box) & \xmark & SR \\
\cite{kalantar2024grid} & \xmark & NS & P, Q & \cmark (PQ Box) & \xmark & Iterative \\
\cite{bandeira2024adp} & \xmark & \xmark & P, Q, V & \cmark (Polyhedron) & \cmark & NS \\
\cite{sarstedt2022monetarization} & \cmark & \xmark & P, Q & \xmark & \xmark & NS \\
\cite{churkin2023impacts} & \cmark & \cmark & P, Q & \xmark & \xmark & NS \\
\cite{capitanescu2023computing} & \cmark & \xmark & P or Q & \xmark & \cmark & NS \\
\cite{bozionek2022design} & \cmark & NS & Q & \cmark (1D Q-Range) & \cmark & SR \\
\textbf{\begin{tabular}[c]{@{}c@{}}This Work\end{tabular}} & \cmark & \cmark & P, Q, V & \cmark (Imp. Polynomial) & \cmark & SR \\ \bottomrule 
\multicolumn{4}{l}{NS: Not Specified, SR: Single Round}
\end{tabular}%
}
\vspace{-0.5cm}
\end{table}

\vspace{-0.4 cm}

\subsection{Contributions}

Considering all the aforementioned challenges, in the present paper, we first determine the three-dimensional PQV FOR using a novel OPF-based approach within the AC-OPF formulation, explicitly accounting for voltage variability. To achieve this, we develop bounding box projection and Fibonacci direction sampling techniques that allow effective sampling on the boundary of the FOR with a relatively small dataset. Additionally, we incorporate diverse FPU characteristics, moving beyond the conventional assumption of ideal rectangular PQ capability profiles. Subsequently, we represent the FOR analytically using an implicit polynomial fitting approach. This method enables the FOR to be expressed as a polynomial in terms of P, Q and V. As a result, the derived polynomial representation can be efficiently integrated into the TSO’s OPF problem. Importantly, the polynomial is constructed conservatively—neither overly restrictive nor excessively permissive—ensuring that all OPF solutions remain feasible without significant loss of available flexibility. Finally, we monetize the FOR by fitting a quadratic function over the PQV domain. This cost representation captures the economic value of aggregated DS flexibility and can be seamlessly incorporated into the OPF formulation. To demonstrate the effectiveness of the proposed method in leveraging DS flexibility for power system management, we conduct a comprehensive comparison against the standard AC-OPF formulation.

The key contributions of the present paper are as follows:
\begin{itemize}
\item A novel and accurate characterization of DS flexibility is proposed that moves beyond conventional PQ-only models. This is achieved by defining the flexibility in the three-dimensional PQV domain, which involves modeling both the FOR boundary and the cost of operating point.
\item A complete analytical package for DS flexibility is formulated, comprising two key components that facilitate its direct and seamless integration into the TSO’s AC-OPF: (i) an implicit polynomial to represent the non-convex FOR boundary, and (ii) a quadratic function to model the cost of operating point.
\item A data-driven methodology is developed to construct the analytical models through novel, tailored AC-OPF-based effective sampling strategies. A complementary approach, combining bounding box projection and Fibonacci directions, is used to efficiently capture the FOR boundary, while Latin hypercube sampling generates a homogeneous dataset for the cost function.
\item A single-round TSO-DSO coordination framework is established, eliminating the need for iterative coordination and computationally intensive disaggregation. The framework is inherently privacy-preserving, exchanging only analytical functions defined over non-sensitive coupling variables, thus obviating the need for the DSO to disclose sensitive data such as network topology and load profiles.
\end{itemize}

The rest of the paper is organized as follows: In Section \ref{sec:method}, we introduce the proposed methodology. In Section \ref{sec:sampling}, we present the sampling strategies employed to generate representative data from the FOR. Subsequently, in Section \ref{sec:construction}, we detail the construction of the analytical functions or the FOR and cost functions. Then, we conduct the case studies in Section \ref{sec:case}. Finally, we provide our conclusions in Section \ref{sec:conclusion}.

\vspace{-0.3 cm}

\section{Overview of the Proposed Methodology}\label{sec:method}

Throughout this paper, we adopt the following conventions: parameters are denoted by standard letters ($a, A$), while variables are represented using boldface letters ($\boldsymbol{a}$, $\boldsymbol{A}$). Sets are expressed using calligraphic letters ($\mathcal{A}$). Matrices appear in uppercase letters ($A$), whereas scalars and (column) vectors are denoted in lowercase ($a$). Functions are written in the form $A(\cdot)$. For a vector $a$, the $n$-th element is written as $a^{(n)}$, while for a matrix $A$, $A^{(n,:)}$ refers to the $n$-th row, and $A^{(i, j)}$ denotes the element in the $i$-th row and $j$-th column. Furthermore, $\preceq$ and $\succeq$ are used for element-wise comparisons, $\leq$ and $\geq$ denote standard scalar comparisons.

To distinguish between system levels, variables pertaining to the integrated transmission-distribution system are denoted with a hat ($\boldsymbol{\widehat{a}}$), those associated only with the TS are marked with an inverted hat ($\boldsymbol{\widecheck{a}}$), and variables specific to the DS are left unmarked ($\boldsymbol{a}$). For instance, $\boldsymbol{\widehat{v}}$, represents the voltage magnitudes at all buses in the integrated system, while $\boldsymbol{\widecheck{v}}$ refers exclusively to voltages at TS buses.

While the presented framework is applicable to any type of FPUs, for ease of exposition, we refer to these units simply as DGs throughout the remainder of the paper.

\vspace{-0.3 cm}

\subsection{Standard AC-OPF for Integrated Transmission-Distribution System}\label{sec:std_opf}

For an integrated transmission-distribution system with $n_{\mathrm{b}}$ total buses, including $n_{\mathrm{g}}$ conventional generator and $n_{\mathrm{b,ts}}$ buses in TS, and comprising $n_{\mathrm{ds}}$ DSs, where the $j$-th DS contains $n_{\mathrm{dg},j}$ DGs, the standard AC-OPF can be formulated as follows:

\begin{subequations}\label{eq:opf_dsotso}
\begin{align}
\min_{\substack{\boldsymbol{\widehat{v}},\boldsymbol{\widehat{\theta}}, \\ \boldsymbol{\widecheck{p}}_{\mathrm{g}}, \boldsymbol{\widecheck{q}}_{\mathrm{g}}, \\ \boldsymbol{p}_{\mathrm{dg},j}, \\ \boldsymbol{q}_{\mathrm{dg},j}}} \quad & \sum_{i = 1}^{n_{\mathrm{g}}}C_{i}(\boldsymbol{{\widecheck{p}}}_{\mathrm{g}}^{(i)}) + \sum_{j = 1}^{n_{\mathrm{ds}}}\sum_{k = 1}^{n_{\mathrm{dg,}j}}C_{jk}(\boldsymbol{p}_{\mathrm{dg,}j}^{(k)}) \label{eq:opf_dsotso0}\\
\text{s.t.} \quad & {G}_{\mathrm{P}}(\boldsymbol{\widehat{v}},  \boldsymbol{\widehat{\theta}}; \widehat{Y}) + \widehat{p}_{\mathrm{d}} - K \boldsymbol{\widecheck{p}}_{\mathrm{g}} - \sum_{j=1}^{n_{\mathrm{ds}}} H_{j} \boldsymbol{p}_{\mathrm{dg,}j} = 0, \label{eq:opf_dsotso1} \\
& {G}_{\mathrm{Q}}(\boldsymbol{\widehat{v}},  \boldsymbol{\widehat{\theta}}; \widehat{Y}) + \widehat{q}_{\mathrm{d}} - K \boldsymbol{\widecheck{q}}_{\mathrm{g}} - \sum_{j=1}^{n_{\mathrm{ds}}} H_{j} \boldsymbol{q}_{\mathrm{dg,}j} = 0, \label{eq:opf_dsotso2} \\
& G_{\mathrm{line}}(\boldsymbol{\widehat{v}},  \boldsymbol{\widehat{\theta}}; \widehat{Y}) \preceq \widehat{l}_{\mathrm{line, max}}, \label{eq:opf_dsotso3} \\
& G_{\mathrm{dg}, j}(\boldsymbol{p}_{\mathrm{dg}, j}, \boldsymbol{q}_{\mathrm{dg}, j}) \preceq 0, \ \forall j \in \{1,.., n_{\mathrm{ds}}\}, \label{eq:opf_dsotso3a}\\
& \widehat{v}_{\mathrm{min}} \preceq \boldsymbol{\widehat{v}} \preceq \widehat{v}_{\mathrm{max}}, \ \widehat{\theta}_{\mathrm{min}} \preceq \boldsymbol{\widehat{\theta}} \preceq \widehat{\theta}_{\mathrm{max}}, \label{eq:opf_dsotso4} \\
& \widecheck{p}_{\mathrm{g, min}} \preceq \boldsymbol{\widecheck{p}}_{\mathrm{g}} \preceq \widecheck{p}_{\mathrm{g, max}}, \ \widecheck{q}_{\mathrm{g, min}} \preceq \boldsymbol{\widecheck{q}}_{\mathrm{g}} \preceq \widecheck{q}_{\mathrm{g, max}}, \label{eq:opf_dsotso5} \\
& p_{\mathrm{dg},j, \mathrm{min}} \preceq \boldsymbol{p}_{\mathrm{dg}, j} \preceq p_{\mathrm{dg},j, \mathrm{max}}, \ \forall j \in \{1,., n_{\mathrm{ds}}\}, \label{eq:opf_dsotso6} \\
& q_{\mathrm{dg},j, \mathrm{min}} \preceq \boldsymbol{q}_{\mathrm{dg}, j} \preceq q_{\mathrm{dg},j, \mathrm{max}}, \ \forall j \in \{1,.., n_{\mathrm{ds}}\}, \label{eq:opf_dsotso7}
\end{align}
\end{subequations}

\noindent where $\boldsymbol{\widehat{v}}, \boldsymbol{\widehat{\theta}}, \widehat{p}_{\mathrm{d}}, \widehat{q}_{\mathrm{d}} \in \mathbb{R}^{n_{\mathrm{b}}}$ denote the bus voltage magnitudes, voltage angles, and active and reactive power demand vectors, respectively. The bus admittance matrix is represented by $\widehat{Y} \in \mathbb{R}^{{n}_{\mathrm{b}} \times {n}_{\mathrm{b}}}$. Active and reactive power generation vectors for the TS are $\boldsymbol{\widecheck{p}}_{\mathrm{g}}, \boldsymbol{\widecheck{q}}_{\mathrm{g}} \in \mathbb{R}^{n_{\mathrm{b,ts}}}$, with the corresponding connection matrix $K \in \mathbb{R}^{n_{\mathrm{b}} \times n_{\mathrm{b,ts}}}$, where $K^{(t,v)} = 1$ if this element is in the TS, and zero otherwise. For the $j$-th DS, $\boldsymbol{p}_{\mathrm{dg},j}, \boldsymbol{q}_{\mathrm{dg},j} \in \mathbb{R}^{n_\mathrm{dg},j}$ denote the active and reactive power generation vectors of DGs. Their connection to the network is captured by matrix $H_{j} \in \mathbb{R}^{n_{\mathrm{b}} \times n_{\mathrm{dg},j}}$ with $H_{j}^{(m,n)} = 1$ if the $n$-th DG of DS $j$ connects at bus $m$, and zero otherwise.

The OPF objective in \eqref{eq:opf_dsotso0} minimizes the total generation cost, including that of DGs. Let $C_{i}(\cdot)$ and $C_{jk}(\cdot)$ be the generation cost functions for TS generators at bus $i$, and the $k$-th DG in DS $j$, respectively. Both functions are modeled as standard quadratic cost functions as $C_{l}(\boldsymbol{p}) = a_{l}\boldsymbol{p}^{2}+b_{l}\boldsymbol{p}+c_{l}$, without loss of generality. For notational simplicity, the first $n_{\mathrm{g}}$ buses are correspond to conventional generators in the integrated system. Equations \eqref{eq:opf_dsotso1}–\eqref{eq:opf_dsotso2} enforce active and reactive power balance through functions $G_{P}(\cdot)$ and $G_{Q}(\cdot)$. The line flow limits are enforced in \eqref{eq:opf_dsotso3} via $G_{line}(\cdot)$, bounded by line flow limit vector $\widehat{l}_{\mathrm{line, max}}$. Additionally, the function $G_{\mathrm{dg}, j}(\cdot)$ in \eqref{eq:opf_dsotso3a}  characterizes the operating limits of DGs that do not exhibit conventional rectangular PQ capability curves. Constraints on voltages and generations are imposed in \eqref{eq:opf_dsotso4}–\eqref{eq:opf_dsotso7}.

\subsection{FOR-Based AC-OPF for Privacy-Preserving DS Flexibility Utilization}\label{sec:pro_opf}

A key observation from \eqref{eq:opf_dsotso} is that utilizing DS flexibility within an OPF framework typically necessitates access to sensitive system information. For instance, the admittance matrix $\widehat{Y}$ encodes detailed grid topology, while the vectors $\widehat{p}_{\mathrm{d}}$ and $\widehat{q}_{\mathrm{d}}$ contain detailed load profiles. As the OPF problem is typically coordinated by the TSO, DSOs are often reluctant to disclose such data due to privacy concerns. To overcome this limitation, we propose a novel FOR-based AC-OPF formulation that enables the integration of DS flexibility without requiring the exchange of sensitive information between TSOs and DSOs.

In the proposed framework, each DS is represented via two analytical functions: one representing its FOR and the other modeling the associated cost of FOR. These functions are defined over non-sensitive coupling variables that are already exchanged between TSOs and DSOs. Specifically, we consider $n_{\mathrm{ds}}$ DSs, where the $j$-th DS comprises $n_{\mathrm{dg},j}$ DGs, and is connected to a designated transmission buses $tb_{j}$, which serve as points of common coupling (PCCs). These PCCs are assumed to be \textit{empty buses}, meaning they do not host any directly connected generation or load units. Each DS $j$ is therefore characterized by functions defined over the coupling variables in the PQV domain, $\boldsymbol{p}_{j}, \boldsymbol{q}_{j}, \boldsymbol{v}_{j} \in \mathbb{R}$, namely active and reactive power flow at the PCC-directed from DS towards TS-and voltage magnitude at the corresponding PCC. For notational convenience, we define a concatenated vector of coupling variables as $\boldsymbol{x}_{j} = {\begin{bmatrix}
 {\boldsymbol{p}_{j}} &{\boldsymbol{q}_{j}}&{\boldsymbol{v}_{j}}\end{bmatrix}}^{\top} \in \mathbb{R}^{3}$. With this, the FOR-based AC-OPF problem can be expressed as follows:

\begin{subequations}\label{eq:opf}
\begin{align}
\min_{\substack{\boldsymbol{\widecheck{v}},\boldsymbol{\widecheck{\theta}}, \\ \boldsymbol{\widecheck{p}}_{\mathrm{g}}, \boldsymbol{\widecheck{q}}_{\mathrm{g}}, \\ \ \boldsymbol{x}_{j}}},  \quad & \sum_{i = 1}^{n_{\mathrm{g}}}C_{i}(\boldsymbol{{\widecheck{p}}}_{\mathrm{g}}^{(i)}) + \sum_{j = 1}^{n_{\mathrm{ds}}}C_{j}(\boldsymbol{x}_{j}) \label{eq:opf0}\\
\text{s.t.} \quad 
& {G}_{\mathrm{P}}(\boldsymbol{\widecheck{v}},  \boldsymbol{\widecheck{\theta}}; \widecheck{Y}) + \widecheck{p}_{\mathrm{d}} - \boldsymbol{\widecheck{p}}_{\mathrm{g}} = 0, \label{eq:opf1} \\
& {G}_{\mathrm{Q}}(\boldsymbol{\widecheck{v}},  \boldsymbol{\widecheck{\theta}}; \widecheck{Y}) + \widecheck{q}_{\mathrm{d}} - \boldsymbol{\widecheck{q}}_{\mathrm{g}} = 0, \label{eq:opf2} \\
& G_{\mathrm{line}}(\boldsymbol{\widecheck{v}},  \boldsymbol{\widecheck{\theta}}; \widecheck{Y}) \preceq \widecheck{l}_{\mathrm{line, max}}, \label{eq:opf3} \\
& \widecheck{v}_{\mathrm{min}} \preceq \boldsymbol{\widecheck{v}} \preceq \widecheck{v}_{\mathrm{max}}, \ \widecheck{\theta}_{\mathrm{min}} \preceq \boldsymbol{\widecheck{\theta}} \preceq \widecheck{\theta}_{\mathrm{max}}, \label{eq:opf4} \\
& \widecheck{p}_{\mathrm{g, min}} \preceq \boldsymbol{\widecheck{p}}_{\mathrm{g}} \preceq \widecheck{p}_{\mathrm{g, max}}, \ \widecheck{q}_{\mathrm{g, min}} \preceq \boldsymbol{\widecheck{q}}_{\mathrm{g}} \preceq \widecheck{q}_{\mathrm{g, max}}, \label{eq:opf5}\\
& FOR_{j}(\boldsymbol{x}_{j}) \leq 0, \ \forall j \in \{1,.., n_{\mathrm{ds}}\}, \label{eq:opf6}\\
& \boldsymbol{\widecheck{p}}_{\mathrm{g}}^{(tb_{j})} = \boldsymbol{p}_{j}, \ \boldsymbol{\widecheck{q}}_{\mathrm{g}}^{(tb_{j})} = \boldsymbol{q}_{j}, \ \boldsymbol{\widecheck{v}}^{(tb_{j})} = \boldsymbol{v}_{j}, \label{eq:opf7}\\
& x_{j, \mathrm{min}} \preceq \boldsymbol{x}_{j} \preceq x_{j, \mathrm{max}}, \ \forall j \in \{1,.., n_{\mathrm{ds}}\}, \label{eq:opf8}\\
& \boldsymbol{x}_{j} = {\begin{bmatrix}
 {\boldsymbol{p}_{j}} \  {\boldsymbol{q}_{j}} \ \boldsymbol{v}_{j} \end{bmatrix}}^{\top}, \forall j \in \{1,.., n_{\mathrm{ds}}\}. \label{eq:opf9}
\end{align}
\end{subequations}

Examining \eqref{eq:opf1} - \eqref{eq:opf5}, it is evident that only TS-related variables are explicitly included, while DS-related variables are encapsulated within the functions $FOR_{j}(\boldsymbol{x}_{j})$, as defined in \eqref{eq:opf6}. $FOR_{j}(\boldsymbol{x}_{j})$ characterize the FOR of the DSs, ensuring compliance with internal technical constraints such as voltage and line flow limits. Specifically, $FOR_{j}(\boldsymbol{x}_{j}) \leq 0$ holds if and only if $\boldsymbol{x}_{j}$ lies within the FOR. Otherwise, it indicates a violation of DS constraints. Additionally, the function $C_{j}(\boldsymbol{x}_{j})$ represent the cost associated with the FOR. Furthermore, \eqref{eq:opf7} enforces the physical coupling between TS and DS by ensuring that the coupling variables match at the PCCs. Lastly, \eqref{eq:opf8} defines the bounds on the coupling variables.

\begin{figure}
\centering
\includegraphics[width=0.7\columnwidth]{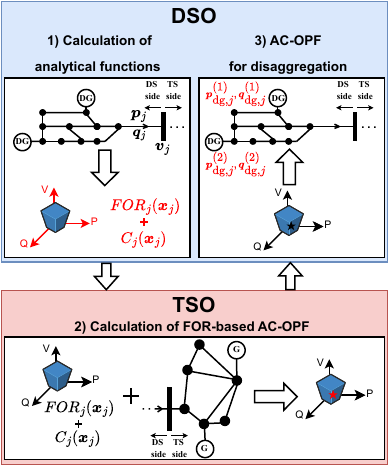}
\caption{Schematic representation of the proposed method. For clarity, only a single DS is illustrated; however, the framework supports multiple DSs. The results obtained at each step are highlighted in red.}
\label{fig:method}
\vspace{-0.5cm}
\end{figure}

Both $FOR_{j}(\cdot)$ and $C_{j}(\cdot)$ must be represented in analytical form to ensure their tractability within the AC-OPF framework. Crucially, these functions are constructed solely using non-sensitive coupling variables, thereby preserving data privacy between TSOs and DSOs. A schematic overview of the proposed framework is illustrated in Fig. \ref{fig:method}. In this architecture, each DSO computes its respective $FOR_{j}(\cdot)$ and $C_{j}(\cdot)$ and communicates them to the TSO. The TSO subsequently incorporates these functions into the FOR-based AC-OPF in \eqref{eq:opf}. Once an optimal solution within the feasible region is obtained, each DSO independently solves its local AC-OPF to determine the internal dispatch of its DGs (or more broadly, FPUs). This process eliminates the need for computationally intensive disaggregation or iterative coordination. As a result, the proposed method enables cost-effective integration of DS flexibility into system-level decision-making, while simultaneously maintaining data confidentiality and complying with the technical constraints of both TSs and DSs.

\section{Sampling Strategies for Flexibility Modeling}\label{sec:sampling}

Constructing the analytical functions $FOR_{j}(\cdot)$ and $C_{j}(\cdot)$ requires two distinct datasets, each tailored to a specific modeling objective. To accurately define the FOR boundary, we generate a dataset by densely sampling points along its surface. In contrast, modeling the cost functions $C_{j}(\cdot)$ requires a homogeneous dataset from the FOR's interior to capture its cost characteristics accurately. We employ efficient and computationally tractable strategies to generate both datasets with minimal overhead, as detailed in the following subsections.

\subsection{Boundary Sampling for the PQV-FOR}
\label{sec:ac-opf-based}

As highlighted in the literature see Section \ref{sec:related_work}), AC-OPF-based methods are particularly effective for identifying the FOR boundary. However, capturing the FOR's complete geometry, which includes both the edges and facets, requires a comprehensive strategy. To this end, we develop an approach for generating FOR boundary data that combines two complementary methods: \textit{Bounding Box Projection Sampling (BBPS)} to effectively capture the edges, and \textit{Fibonacci Direction Sampling (FDS)} to ensure full coverage of the facets. The combination of these methods, which are detailed in the following subsections, yields a well-distributed and representative dataset that accurately characterizes the FOR's complex shape.

\begin{figure*}
\centering
\includegraphics[width=0.7\textwidth]{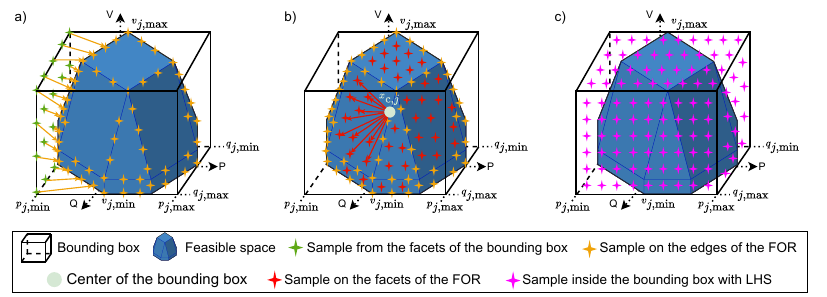}
\caption{The sampling procedure with a) BBPS b) FDS c) LHS. For clarity in the illustration, arrows are indicated only one side.}
\label{fig:sampling}
\vspace{-0.4cm}
\end{figure*}

\subsubsection{Bounding Box Projection Sampling (BBPS)}\label{sec:projection}

The BBPS method, detailed in Algorithm \ref{alg-projection}, is the first of two complementary strategies. The process is executed in two phases.

In the first phase, we establish a tight bounding box that encloses the FOR. This is achieved by solving six AC-OPF problems to find the minimum and maximum feasible values for each of the three coupling variables ($\boldsymbol{p}_j, \boldsymbol{q}_j, \boldsymbol{v}_j$) at the PCC, resulting in the limit vectors $x_{j, \mathrm{min}}, x_{j, \mathrm{max}} \in \mathbb{R}^{3}$. This targeted approach avoids inefficient sampling of the inherently large infeasible space and focuses the data generation effort on the most relevant region.

\begin{algorithm}[H]
    \caption{The BBPS-based Algorithm for Generating FOR Boundary Data}\label{alg-projection}
    
    \begin{algorithmic}[1]
    \item[] \textbf{Input:} Power system data for DS $j$
    \item[] \textbf{Output:} $x_{j, \mathrm{min}}, x_{j, \mathrm{max}}$, \ $D_{\mathrm{bbps}, j}$
    \vspace{0.1cm}
    \State{\textit{Phase 1: Determine Bounding Box}}
    \For{$idx \leftarrow 1$ to $3$}
        \State{$x_{j, \mathrm{min}}^{(idx)} \leftarrow$ Solve $\min \boldsymbol{x}_j^{(idx)}$ s.t. AC-OPF constraints}
        \State{$x_{j, \mathrm{max}}^{(idx)} \leftarrow$ Solve $\max \boldsymbol{x}_j^{(idx)}$ s.t. AC-OPF constraints}
    \EndFor
    \vspace{0.1cm}
    \State{\textit{Phase 2: Project Samples onto FOR Boundary}}
    \State{$D_{\mathrm{bbps}, j} \leftarrow [\,];$}
    \For{$k \leftarrow 1$ to $n_{\mathrm{bbps}, j}$}
        \State{$x_{\mathrm{lhsf},j} \leftarrow$ sample a vector from the facets of the bounding box defined by $x_{j, \mathrm{min}}$ and $x_{j, \mathrm{max}}$ using LHS}
        \State{Define optimization problem:}
        \begin{subequations}\label{eq:projection}
        \begin{align}
        \min_{\boldsymbol{x}_j} \ &  \left\|x_{\mathrm{lhsf},j} - \boldsymbol{x}_{j} \right\|_{2}^{2} \\
        \text{s.t.} \ & \text{Standard AC-OPF constraints for DS } j \nonumber
        \end{align}
        \end{subequations}
        \State{Solve \eqref{eq:projection} and obtain optimal point $x_{j}^{*}$}
        \State{$D_{\mathrm{bbps}, j} \leftarrow [D_{\mathrm{bbps}, j}; {x_{j}^{*}}^{\top}];$}
    \EndFor
    \end{algorithmic}
\end{algorithm}

In the second phase, we generate the boundary dataset itself. A set of $n_{\mathrm{bbps}, j}$ points, denoted individually as $x_{\mathrm{lhsf}, j} \in \mathbb{R}^{3}$, is sampled on the facets of the previously determined bounding box using Latin Hypercube Sampling (LHS) \cite{huntington1998improvements}. Each external point $x_{\mathrm{lhsf}, j}$ is then projected onto the FOR boundary by solving the optimization problem defined in \eqref{eq:projection}. This problem finds the closest feasible point $x_{j}^{*} \in \mathbb{R}^{3}$ on the FOR by minimizing the Euclidean ($L_2$) distance, an approach similar to the one introduced in our previous work \cite{dindar2025machine}.

The resulting optimal points $x_j^*$ are collected to form the dataset $D_{\mathrm{bbps}, j} \in \mathbb{R}^{n_{\mathrm{bbps}, j} \times 3}$. The points generated via BBPS tend to concentrate along the sharp edges of the FOR (see Fig. \ref{fig:sampling}a), effectively capturing these critical features.

\subsubsection{Fibonacci Direction Sampling (FDS)}\label{sec:arrow}

While the BBPS method effectively captures the edges of the FOR, a complementary approach is needed to sample its facets. To this end, we introduce the FDS method, detailed in Algorithm \ref{alg-arrow}, which ensures comprehensive coverage of the FOR's entire surface (see Fig. \ref{fig:sampling}b).

\begin{algorithm}[H]
    \caption{The FDS-based Algorithm for Generating FOR Boundary Data}\label{alg-arrow}
    
    \begin{algorithmic}[1]
    \item[] \textbf{Input:} Power system data for DS $j$, $x_{j, \mathrm{min}}, x_{j, \mathrm{max}}$
    \item[] \textbf{Output:} $D_{\mathrm{fds}, j}$

    \State{Define an optimization problem for a given direction $d_k$:}
        \begin{subequations}\label{eq:arrow}
        \begin{align}
        \max_{\boldsymbol{t}_k} \ & \boldsymbol{t}_{k} \\
        \text{s.t.} \ &  \boldsymbol{x}_{j} = x_{\mathrm{c}, j} + d_{k}\boldsymbol{t}_{k}  \\
        \ & \text{Standard AC-OPF constraints for DS } j \nonumber
        \end{align}
        \end{subequations}

    \State{$D_{\mathrm{fds}, j} \leftarrow [\,];$}
    \State{$x_{\mathrm{c}, j} \leftarrow 0.5 \times (x_{j, \mathrm{max}} + x_{j, \mathrm{min}})$}
    \State{$\phi \leftarrow \pi (3 - \sqrt{5})$} \Comment{Golden angle for Fibonacci lattice}
    
    \For{$k \leftarrow 0$ to $n_{\mathrm{fds}, j}-1$}
        \State \parbox[t]{\linewidth}{$d_k \leftarrow \begin{bmatrix} r \cos(\theta)\ r \sin(\theta)\ z \end{bmatrix}^\top$, \\
        where $z = 1 - \frac{2k + 1}{n_{\mathrm{fds}, j}}$, \ $\theta = \phi \cdot k$, \ $r = \sqrt{1 - z^2}$}
        
        \State{Solve \eqref{eq:arrow} to obtain the optimal step size $t_{k}^{*}$} 
        \State{Calculate the boundary point: $x_{j}^{*} \leftarrow x_{\mathrm{c}, j} + d_{k}t_{k}^{*}$}
        \State{$D_{\mathrm{fds}, j} \leftarrow [D_{\mathrm{fds},j}; {x_{j}^{*}}^{\top}];$}
    \EndFor
    \end{algorithmic}
\end{algorithm}

The core idea is to cast \textit{virtual arrows} from the FOR's interior to its boundary along systematically chosen directions. The process begins by defining an origin point, $x_{\mathrm{c}, j} \in \mathbb{R}^{3}$, at the center of the bounding box. From this origin, a set of $n_{\mathrm{fds}, j}$ direction vectors, $d_k \in \mathbb{R}^{3}$, is generated using the Fibonacci lattice method, which ensures uniform angular coverage in three-dimensional space. For each direction $d_k$, we solve the optimization problem in \eqref{eq:arrow} to find the maximum feasible step size, $\boldsymbol{t}_{k} \in \mathbb{R}$, along that direction. The resulting intersection point with the FOR boundary, ${x_{j}^{*}} \in \mathbb{R}^3$, is then recorded. This procedure is repeated for all directions, yielding the dataset $D_{\mathrm{fds}, j} \in \mathbb{R}^{n_{\mathrm{fds}, j} \times 3}$.

Finally, to create a complete representation of the FOR's geometry, the datasets from both sampling methods are combined into a single dataset: $D_{\mathrm{FOR}, j} = [D_{\mathrm{bbps}, j}; D_{\mathrm{fds}, j}]$, where, $D_{\mathrm{FOR}, j} \in \mathbb{R}^{n_{\mathrm{FOR}, j} \times 3}$. For this, $n_{\mathrm{FOR}, j}$ is defined as $n_{\mathrm{FOR}, j} = n_{\mathrm{bbps}, j} + n_{\mathrm{fds}, j}$. This final dataset captures both the sharp edges and the smooth facets of the FOR with high fidelity, serving as the foundation for the data-driven construction of the analytical model $FOR(\boldsymbol{x})$.

\subsection{Interior Sampling for the Cost of the FOR}\label{sec:cost}

To create the cost function, a dataset of feasible operating points from the interior of the FOR is required. Unlike the boundary-focused methods, this step necessitates a homogeneous sampling of the entire FOR. We achieve this using the LHS method, as illustrated in Fig. \ref{fig:sampling}c and detailed in Alg. \ref{alg-cost}.

\begin{algorithm}[H]
    \caption{The LHS-based Algorithm for Generating Cost Data}\label{alg-cost}
    
    \begin{algorithmic}[1]
    \item[] \textbf{Input:} Power system data for DS $j$, $x_{j, \mathrm{min}}, x_{j, \mathrm{max}}$
    \item[] \textbf{Output:} $D_{\mathrm{cost}, j}$, $y_{\mathrm{cost}, j}$

    \State{Define an optimization problem for a candidate point $x_{\mathrm{lhsc},j}$:}
        \begin{subequations}\label{eq:cost}
        \begin{align}
        \min \ & C_j = \sum_{k = 1}^{n_{\mathrm{dg,}j}}C_{jk}(\boldsymbol{p}_{\mathrm{dg,}j}^{(k)})  \\
        \text{s.t.} \ & \boldsymbol{x}_{j} = x_{\mathrm{lhsc},j} \\
        \ & \text{Standard AC-OPF constraints for DS } j \nonumber
        \end{align}
        \end{subequations}      

    \State{$D_{\mathrm{cost}, j} \leftarrow [\,];$}
    \State{$y_{\mathrm{cost}, j} \leftarrow [\,];$}
    \State{$idx \leftarrow 1;$}
        
    \While{$idx \leq n_{\mathrm{cost}, j}$}
        \State{$x_{\mathrm{lhsc},j} \leftarrow$ sample a vector from the bounding box using LHS}      
        \If{$\eqref{eq:cost}$ with $x_{\mathrm{lhsc},j}$ is feasible}
            \State{Obtain the optimal total cost $C_j^*$}
            \State{$D_{\mathrm{cost}, j} \leftarrow [D_{\mathrm{cost},j}; {x_{\mathrm{lhsc},j}}^{\top}];$}
            \State{$y_{\mathrm{cost}, j} \leftarrow [y_{\mathrm{cost},j}; C_j^*];$}
            \State{$idx \leftarrow idx + 1; $}
        \EndIf
    \EndWhile
    \end{algorithmic}
\end{algorithm}

The process begins by generating a candidate operating point, $x_{\mathrm{lhsc},j} \in \mathbb{R}^{3}$, from within the bounding box using LHS. This point is then tested for feasibility by solving the AC-OPF problem defined in \eqref{eq:cost}, where the PCC variables are fixed to the candidate's values. If feasible, the candidate point is valid, and the point itself is stored in the feature dataset $D_{\mathrm{cost}, j} \in \mathbb{R}^{n_{\mathrm{cost}, j} \times 3}$, and the corresponding optimal total cost, $C_j^*$, is stored in the target vector $y_{\mathrm{cost}, j} \in \mathbb{R}$. This procedure is repeated until the desired number of samples, $n_{\mathrm{cost}, j}$, is collected, resulting in a dataset that homogeneously covers the cost characteristics of the FOR interior.

\section{Construction of the Analytical Functions}\label{sec:construction}

In the following, we detail the procedure for constructing the analytical functions $FOR_{j}(\cdot)$ and $C_{j}(\cdot)$ based on the datasets $D_{\mathrm{FOR}, j}$ and $D_{\mathrm{cost}, j}$, respectively. For notational brevity, we will drop the DS subscript $j$ for the remainder of this section.

\subsection{Implicit Polynomial Representation of the PQV-FOR}\label{sec:FOR_const}

We approximate the FOR using an implicit polynomial of degree $d_{\mathrm{FOR}}$. A point $x = (p, q, v)$ is considered to be on the boundary of the FOR if it satisfies $FOR(\boldsymbol{x}) = 0$. The polynomial is of the form:
\begin{equation}\label{eq:for}
FOR(\boldsymbol{x}) = \sum_{\alpha+\beta+\theta \leq d_{\mathrm{FOR}}} a_{\alpha\beta\theta} \boldsymbol{p}^{\alpha}\boldsymbol{q}^{\beta}\boldsymbol{v}^{\theta}.
\end{equation}

The coefficients $\boldsymbol{a}_{\alpha\beta\theta}$ are the unknown parameters we seek. To facilitate a linear algebraic solution, we establish an explicit ordering for the monomial terms. We define a bijective mapping $\sigma: \{(\alpha,\beta,\theta) \in \mathbb{N}_0^3 \mid \alpha+\beta+\theta \le d_{\mathrm{FOR}}\} \to \{1, \dots, K\}$ that uniquely maps each exponent triplet $(\alpha,\beta,\theta)$ to an index $s$, where, $K = \binom{d_{\mathrm{FOR}}+3}{3}$ is the total number of monomial terms. This mapping allows the coefficients to be arranged into a single column vector $a \in \mathbb{R}^K$, where the $s$-th element, $\boldsymbol{a}^{(s)}$, corresponds to the coefficient $\boldsymbol{a}_{\alpha\beta\theta}$ such that $\sigma(\alpha,\beta,\theta) = s$.

\subsubsection{Volumetric Constraint Generation}

To ensure the polynomial defines a volume, yielding negative values inside the FOR, positive values outside, and values (near) zero on the boundary, we augment the original boundary data matrix. This volumetric constraint strategy is inspired by the 3L algorithm for fitting implicit surfaces~\cite{blane20023l}.

Let the original data matrix be denoted $D_\mathrm{bnd} \equiv D_\mathrm{FOR}$. We generate additional constraint sets by scaling the boundary points relative to their centroid, $c$, defined as:
\begin{equation}
c = \frac{1}{n_{\mathrm{FOR}}} \sum_{l=1}^{n_{\mathrm{FOR}}} x_l
\end{equation}
where $x_l$ is the $l$-th row (sample point) of $D_\mathrm{bnd}$. From each point $x_{\mathrm{bnd}}$ in the rows of $D_\mathrm{bnd}$, we generate new points by applying a scaling factor $\gamma$. An inner point $x_\mathrm{in}$ is generated using a shrink factor $\gamma_\mathrm{in} < 1$, and one or more sets of outer points $\{x_{\mathrm{out},m}\}$ are generated using growth factors $\gamma_{\mathrm{out},m} > 1$. The transformation is given by:

\begin{equation}
x_\mathrm{new} = c + \gamma (x_\mathrm{bnd} - c).
\end{equation}

This procedure yields an inner data matrix $D_\mathrm{in}$ and a set of $S$ outer data matrices $\{D_{\mathrm{out},m}\}_{m=1}^S$, each of size $n \times 3$.

\subsubsection{Linear System Formulation and Solution}

The problem of finding the coefficient vector $\boldsymbol{a}$ is framed as solving an overdetermined system of linear equations. For each constraint data matrix, we construct a corresponding monomial matrix. Let $D'$ be a generic data matrix.
The monomial matrix $M(D') \in \mathbb{R}^{|D'| \times K}$ is constructed such that each row corresponds to a point $x_l$ (the $l$-th row of $D'$) and each column $s$ corresponds to a unique monomial basis term, ordered according to the mapping $\sigma$.
The entry for the $l$-th row and the $s$-th column is given by the evaluation of the $s$-th monomial at point $x_l$ as $M(D')^{(l,s)} = p_l^{\alpha}q_l^{\beta}v_l^{\theta}$, where  $s=\sigma(\alpha,\beta,\theta)$.

Using this definition, we construct a monomial matrix for each of our constraint sets: $M_\mathrm{in} = M(D_\mathrm{in})$, $M_\mathrm{bnd} = M(D_\mathrm{bnd})$, and $M_{\mathrm{out},m} = M(D_{\mathrm{out},m})$ for $m=1, \dots, S$. These matrices are vertically concatenated to form a single composite system matrix $M_{\mathrm{FOR}}$. Correspondingly, we define a constraint vector $b_{\mathrm{FOR}}$ by concatenating vectors of target values for each region. These target vectors consist of small constants chosen to enforce the desired sign of the polynomial: $c_\mathrm{in} < 0$, $c_\mathrm{bnd} \approx 0$, and $c_{\mathrm{out},m} > 0$. The complete linear system is thus formulated as $M_{\mathrm{FOR}}\boldsymbol{a} = b_{\mathrm{FOR}}$, where
\begin{equation}
M_{\textrm{FOR}} =
\begin{bmatrix}
M_\mathrm{in} \\
M_\mathrm{bnd} \\
M_{\mathrm{out},1} \\
\vdots \\
M_{\mathrm{out},S}
\end{bmatrix}
, \
b_{\textrm{FOR}} =
\begin{bmatrix}
c_\mathrm{in} \cdot \mathrm{1} \\
c_\mathrm{bnd} \cdot \mathrm{1} \\
c_{\mathrm{out},1} \cdot \mathrm{1} \\
\vdots \\
c_{\mathrm{out},S} \cdot \mathrm{1}
\end{bmatrix}.
\end{equation}

Thereby, $\mathrm{1}$ is a column vector of ones of appropriate dimension. Since this system is overdetermined, an exact solution generally does not exist. We therefore seek the coefficient vector $a^*$ that minimizes the squared Euclidean norm of the residual, $\|M\boldsymbol{a} - b\|^2$. This least-squares solution is found using the Moore-Penrose pseudoinverse, denoted by $M_{\textrm{FOR}}^\dagger$:
\begin{equation}
a^* = M_{\textrm{FOR}}^\dagger b_{\textrm{FOR}}.
\end{equation}

The resulting vector $a^*$ contains the coefficients of the implicit polynomial $FOR(\boldsymbol{x})$ in \eqref{eq:for}, providing an analytical representation of the FOR.

\begin{figure*}
\centering
\includegraphics[width=\textwidth]{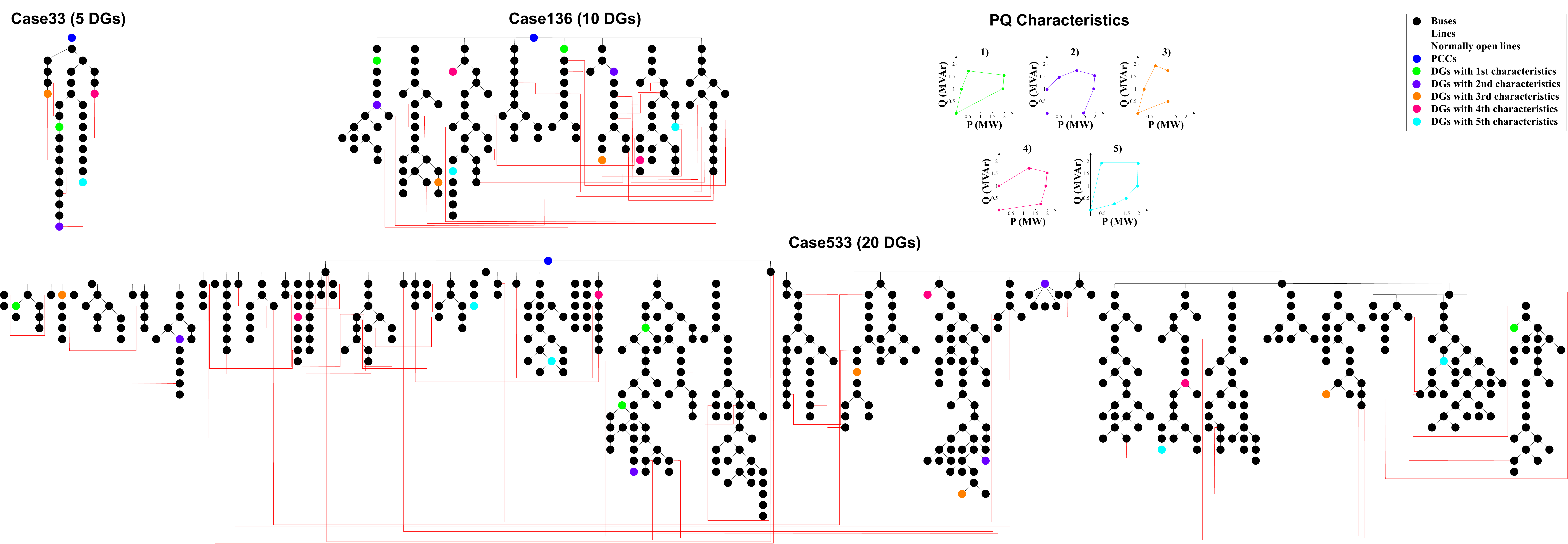}
\caption{Single line diagrams of the DSs and characteristics of the DGs.}
\label{fig:DSs}
\vspace{-0.4cm}
\end{figure*}

\subsection{Quadratic Representation of the Cost of the FOR}
\label{sec:cost_const}

To monetize the FOR, we model the cost function $C(\boldsymbol{x})$ as a trivariate quadratic polynomial. This is a special case of the polynomial fitting described in the preceding section, with the degree fixed to $d_{\mathrm{cost}}=2$. The cost function takes the general form:

\begin{equation}\label{eq:C}
    C(\boldsymbol{x}) = \sum_{\alpha+\beta+\theta \le 2} w_{\alpha\beta\theta} \boldsymbol{p}^\alpha \boldsymbol{q}^\beta \boldsymbol{v}^\theta.
\end{equation}
The fitting procedure seeks the coefficients $\boldsymbol{w}_{\alpha\beta\theta}$ that best match the data generated by Algorithm \ref{alg-cost}. It follows the same linear least-squares formulation previously described. The monomial matrix $M_{\mathrm{cost}} \in \mathbb{R}^{n_{\mathrm{cost}} \times K}$ is constructed from the feature dataset $D_{\mathrm{cost}}$, where $K=10$ for a second-degree trivariate polynomial. The vector $y_{\mathrm{cost}}$ serves directly as the target for the regression. The coefficient vector $w^* \in \mathbb{R}^{10}$, which contains the ordered coefficients $\boldsymbol{w}_{\alpha\beta\theta}$, is then found by solving the overdetermined system $M_{\mathrm{cost}} \boldsymbol{w} = y_{\mathrm{cost}}$ using the Moore-Penrose pseudoinverse:
\begin{equation}
    w^* = M_{\mathrm{cost}}^\dagger y_{\mathrm{cost}}.
\end{equation}

The resulting vector $w^*$ provides the coefficients for the analytical cost function $C(\boldsymbol{x})$ in \eqref{eq:C}.

\section{Case Studies and Discussion}\label{sec:case}

We assess the performance of the proposed method by comparing it against the traditional AC-OPF, which does not incorporate data privacy considerations. To this end, we first consider different DSs with diverse FPU characteristics. We then approximate the FORs and their associated cost using analytical functions and evaluate the accuracy of these approximations for each DS. Finally, we integrate the analytically approximated DSs into specified TSs, and the proposed FOR-based AC-OPF framework is applied. The performance of this framework is evaluated through extensive case studies, focusing on feasibility, optimality, and computational efficiency.

All simulations are conducted in the \textsc{MATLAB} environment, utilizing the \textsc{MATPOWER} toolbox \cite{MATPOWER} with the \textsc{KNITRO} solver \cite{byrd2006k} for the AC-OPF problems. The case studies are conducted on a PC equipped with an Intel Core i7-10700K CPU @ 3.80 GHz and 32 GB RAM.

\vspace{-0.3cm}

\subsection{Distribution Systems Specifications}\label{sec:case_ds}

To rigorously evaluate the performance and scalability of the proposed method, we conduct case studies on 33-, 136-, and 533-bus DSs (i.e., $n_{\mathrm{ds}}=3$). These DSs are augmented with 5, 10, and 20 DGs, respectively. Notably, the voltage at all PCCs is allowed to vary within a range of 0.95 to 1.05 p.u., rather than being fixed to a nominal value as commonly assumed in the literature. The configurations and key specifications of these systems are illustrated in Fig. \ref{fig:DSs}.

As depicted in Fig. \ref{fig:DSs}, all normally open lines in the selected DSs are closed, enabling meshed network topologies. This allows the proposed method to be tested under more complex meshed configurations, in contrast to many existing studies that focus solely on radial networks. Moreover, instead of assuming idealized rectangular PQ characteristics, each DS is equipped with diverse non-ideal convex PQ characteristics, in line with the modeling approach presented in \cite{dindar2025privacy}. Specifically, five distinct types of DG characteristics are incorporated to reflect the heterogeneity of FPUs. This results in complex, three-dimensional FORs varying with voltage, allowing a comprehensive assessment of the proposed methodology.

\subsection{Derivation of Analytical FOR and Cost Functions for DSs}\label{sec:case_derivation}

\begin{figure*}
\centering
\includegraphics[width=0.85\textwidth]{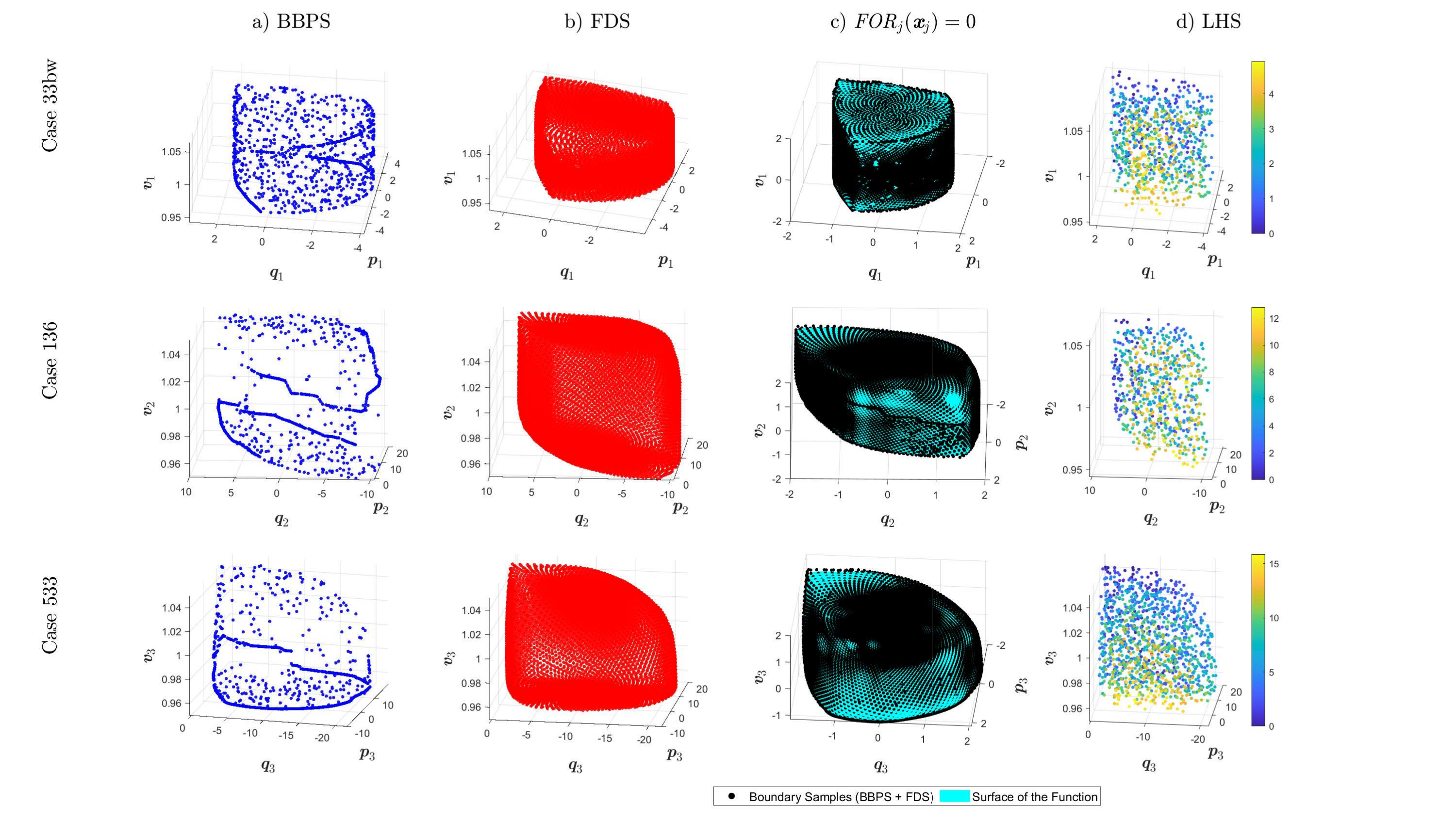}
\caption{a) Dataset generated using BBPS b) Dataset generated using FDS c) Constructed FOR functions (with z-score normalization) d) Cost distribution obtained via LHS. Active and reactive power are expressed in MW and MVAR, respectively, while voltage is expressed in p.u.}
\label{fig:case_study_b}
\vspace{-0.4cm}
\end{figure*}

In this subsection, we construct analytical approximations of the FOR and the corresponding cost functions for each DS. These approximations are derived from the datasets $D_{\mathrm{bbps}, j}$, $D_{\mathrm{fds}, j}$, i.e., $D_{\mathrm{FOR}, j}$ and $D_{\mathrm{cost}, j}$, which are obtained through the tailored sampling strategies described earlier in Algorithms \ref{alg-projection}, \ref{alg-arrow}, \ref{alg-cost}. The number of generated data points for each DS is summarized in Table \ref{tab:parameters}. Note that, due to the inclusion of voltage as an additional variable, constructing the three-dimensional FOR requires relatively more data points compared to its two-dimensional counterpart. Furthermore, the datasets are visualized in the three-dimensional PQV domain in Fig. \ref{fig:case_study_b}a and \ref{fig:case_study_b}b. As observed, the FOR and its associated cost characteristics can be effectively captured. As expected, the BBPS method tends to populate points along the edges of the FOR, whereas the FDS method generates samples that lie predominantly on its facets. Merging these datasets yields an effective representation of the FOR. The visualizations also reveal the emergence of complex FOR geometries, due to meshed topologies and the presence of a large number of DGs.

\begin{table}
\centering
\caption{Parameters for the Design of the Functions}
\label{tab:parameters}
\resizebox{\columnwidth}{!}{%
\begin{tabular}{lcccccccccccc}
\hline
\multirow{2}{*}{DS} & \multirow{2}{*}{$n_{\mathrm{bbps}}$} & \multirow{2}{*}{$n_{\mathrm{fds}}$} & \multirow{2}{*}{$n_{\mathrm{cost}}$} & \multirow{2}{*}{$d_{\mathrm{FOR}}$} & \multirow{2}{*}{$d_{\mathrm{cost}}$} & \multirow{2}{*}{$\gamma_\mathrm{in}$} & \multirow{2}{*}{$\gamma_{\mathrm{out},1}$} & \multirow{2}{*}{$\gamma_{\mathrm{out},2}$} & \multirow{2}{*}{$c_\mathrm{in}$} & \multirow{2}{*}{$c_\mathrm{bnd}$} & \multirow{2}{*}{$c_{\mathrm{out},1}$} & \multirow{2}{*}{$c_{\mathrm{out},2}$}\\
                       &                       &                      \\ \hline
Case 33bw      & 10\textsuperscript{3}             & 10\textsuperscript{4}                    & 10\textsuperscript{3} & 8 & 2 & 0.999 & 1.005 & 1.07 & -0.15 & 0 & 0.1 & 0. 2                  \\ 
Case 136     & 10\textsuperscript{3}       & 10\textsuperscript{4}                    & 10\textsuperscript{3}   & 8 & 2 & 0.999 & 1.02 & 1.03 & -0.05 & 0 & 0.1 & 0. 2                \\ 
Case 533     & 10\textsuperscript{3}       & 10\textsuperscript{4}                    & 10\textsuperscript{3}  & 8 & 2 & 0.999 & 1.005 & 1.07 & -0.08 & 0 & 0.15 & 0. 6                 \\ 
\hline
\end{tabular}%
}
\vspace{-0.5cm}
\end{table}

For the cost modeling, samples are drawn homogeneously from the interior of the FOR using the LHS method (see Fig. \ref{fig:case_study_b}d). The cost function reflects only the generation costs of the DGs; the cost of power imported through the PCC is excluded, as it is already incorporated into the generation costs when the DS model is integrated to the TS. 

After generating the datasets, we  proceed to construct the analytical functions by representing them in polynomial form. The hyperparameters used for this fitting process are summarized in Table \ref{tab:parameters}. Specifically, one inner data matrix $D_\mathrm{in}$ and two outer data matrices $D_\mathrm{out}$ (i.e., $S=2$) are generated. These hyperparameters are deliberately chosen to ensure that the resulting polynomial defines a conservative feasible region. Such conservatism is essential in power system operations: while classifying a feasible operating point as infeasible results in only a small cost penalty, misclassifying an infeasible operating point as feasible can lead to severe operational and reliability issues. The fitted polynomials, shown in Fig. \ref{fig:case_study_b}c together with the boundary data points, demonstrate that the proposed approach achieves a highly accurate approximation of the FOR with minimal conservativeness, ensuring that the approximation remains entirely within the true FOR.

To evaluate the performance of the functions $FOR_{j}(\boldsymbol{x}_{j})$, we generate random 10\textsuperscript{5} samples within the bounding box. These samples naturally contain specific amount of feasible and infeasible samples. Using these samples, we estimate the volume of the FOR that is captured by the polynomial approximation. For this purpose, we adopt the classical confusion matrix framework. The quality of the approximation is then quantified using standard performance metrics derived from the confusion matrix, and the corresponding results are reported in Table \ref{tab:for}.

\begin{table}
\centering
\caption{Performance Metrics of the FOR Functions}
\label{tab:for}
\resizebox{0.8\columnwidth}{!}{%
\begin{tabular}{lcccc}
\hline
\multirow{2}{*}{Function} & \multirow{2}{*}{Fitting time (s)} & \multirow{2}{*}{Accuracy} & \multirow{2}{*}{Recall}  & \multirow{2}{*}{Specificity} \\
                       &                       &                      \\ \hline
$FOR_{1}(\boldsymbol{x}_{1})$      & 0.36             & 99.67\%                 & 99.16\%  &  100\%           \\ 
$FOR_{2}(\boldsymbol{x}_{2})$     & 0.37       & 99.68\%            & 98.97\%       &  100\%            \\ 
$FOR_{3}(\boldsymbol{x}_{3})$     & 0.39       & 99.50\%          & 98.14\%     &  100\%              \\ 
\hline
\end{tabular}%
}
\vspace{-0.5cm}
\end{table}

The specificity metric evaluates the ability of the function to correctly identify infeasible points. Since all functions achieve 100\% specificity, no infeasible points are misclassified as feasible, indicating that the approximation does not allow any infeasible operating points. In contrast, the recall metric measures performance on feasible samples. For example, for $FOR_{1}(\boldsymbol{x}_{1})$,  the recall is 99.16\%, meaning that 99.16\% of feasible points are correctly identified, while 0.84\% of the FOR volume is not captured. This illustrates the trade-off between ensuring complete feasibility and fully representing the volume of the FOR. Overall, when the results are examined, the proposed approximation performs as intended: a small portion of the FOR is sacrificed to reach 100\% feasibility. Note that, the impact of this slight volume loss on the overall operational cost is analyzed in the following section.

Furthermore, the cost functions are constructed using the hyperparameters specified in Table \ref{tab:parameters}, with their performance metrics summarized in Table \ref{tab:cost}. As shown, the cost functions exhibit high accuracy, indicating that they can effectively approximate the cost of the FOR. The corresponding cost distribution is illustrated in Fig. \ref{fig:case_study_b}d. Also, the fitting times of both $FOR_{j}(\boldsymbol{x}_{j})$ and $C_{j}(\boldsymbol{x}_{j})$ are quite low, demonstrating the computational efficiency of the proposed approach. Overall, given that the DSs considered involve a large number of buses and DGs with complex meshed structures, the analytical approximations are obtained with consistently high performance.

\begin{table}
\centering
\caption{Performance Metrics of the Cost Functions}
\label{tab:cost}
\resizebox{0.6\columnwidth}{!}{%
\begin{tabular}{lccc}
\hline
\multirow{2}{*}{Function} & \multirow{2}{*}{Fitting time (s)} & \multirow{2}{*}{RMSE} & \multirow{2}{*}{MAE} \\
                       &                       &                      \\ \hline
$C_{1}(\boldsymbol{x}_{1})$      & 0.041             & 0.0021                    & 0.0016                   \\ 
$C_{2}(\boldsymbol{x}_{2})$     & 0.045       & 0.0055                    & 0.0042                   \\ 
$C_{3}(\boldsymbol{x}_{3})$     & 0.057       & 0.0341                    & 0.0266                   \\ 
\hline
\end{tabular}%
}
\vspace{-0.5cm}
\end{table}

\vspace{-0.3cm}

\subsection{Incorporation of Analytical DS Representations into TS and Benchmarking Against AC-OPF}\label{sec:benchmark}

In this chapter, the approximated DSs are integrated into the TS as formulated in Equation \eqref{eq:opf}. To this end, Case 33bw, Case 136, and Case 533 DSs are integrated with the corresponding TS benchmark models from the PGLib-OPF library, namely Case 30, Case 57, and Case 162, respectively \cite{babaeinejadsarookolaee2019power}. In the Case 30 , buses 11, 16, and 20 are designated as the PCCs; in the Case 57, buses 7, 34, and 48 serve as the PCCs; and in the Case 162, buses 75, 109, and 129 are selected as the PCCs. In this way, three distinct TS–DS test systems are constructed, each consisting of one TS interconnected with three DSs.

\begin{figure}
\centering
\includegraphics[width=\columnwidth]{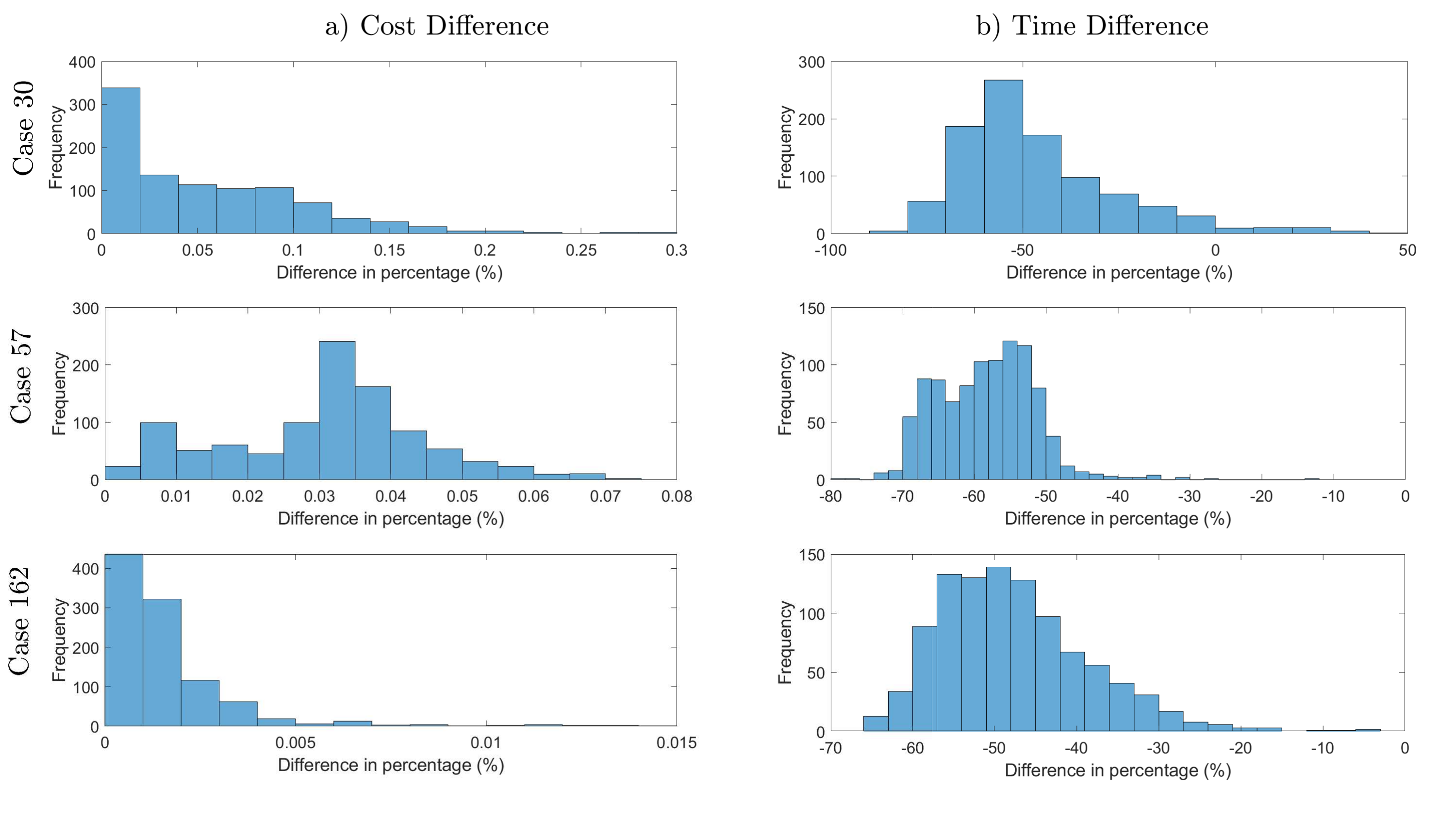}
\caption{Histogram of total cost and computational time differences, with AC-OPF as the reference, for integrated TS-DS. Each TS is combined with three analytically represented DSs (Case 33bw, Case 136, and Case 533).}
\label{fig:histogram}
\vspace{-0.4cm}
\end{figure}

After that, we conduct simulations using 1,000 randomly generated sets of cost coefficients. The results are compared against the standard AC-OPF in terms of both total cost and computational time. Fig. \ref{fig:histogram} presents histograms of the cost and time differences, using the standard AC-OPF as the reference. The feasibility ratio is consistently 100\%, indicating that the proposed method never produces infeasible solutions. The cost difference remains negligible across all cases, with average deviations of 0.058\%, 0.031\%, and 0.002\%, respectively. Although the FOR of the DSs is constructed in a relatively conservative manner, its impact on the total cost is observed to be minimal. As the size of the TS increases, the influence of the DSs diminishes, leading to smaller cost deviations. In terms of computational time, the proposed method outperforms the standard AC-OPF in the majority of tests, demonstrating high computational efficiency. The average time difference are -44.22\%, -58.11\%, -47.62\% (-0.23, -0.24, and -0.33 seconds), respectively. Note that, negative values indicate that the proposed method is faster than the standard AC-OPF. These efficiency gains are achieved by representing the complex DSs with compact polynomial approximations.

Note that, within the proposed method, the TSO determines an operating point located on the FOR by considering the overall system cost. The corresponding cost value at this operating point is already available to the DSO through the previously constructed cost function. Consequently, disaggregation at the DS level can be seamlessly carried out by solving a standard AC-OPF that incorporates this operating point as an input.

Overall, the proposed method achieves 100\% feasibility, negligible cost deviations, and notable computational efficiency, even in the presence of large, meshed DSs with numerous busbars and diverse DG characteristics. This demonstrates the efficiency of the approach. By leveraging analytical polynomial representations, the method enables the tractable integration of DS flexibility into AC-OPF while fully preserving DS-level privacy and adhering to technical constraints.

\vspace{-0.3cm}

\section{Conclusion}\label{sec:conclusion}

In the present paper, we address key challenges in enabling effective coordination between transmission system operators (TSOs) and distribution system operators (DSOs) through a privacy-preserving representation of distribution system (DS) flexibility.  We propose a comprehensive framework for constructing and utilizing a three-dimensional PQV feasible operating region (FOR), that explicitly accounts for voltage variability at the point of common coupling (PCC) and heterogeneous flexibility-providing unit (FPU) characteristics. Our method employs an advanced AC optimal power flow (OPF)-based sampling strategy, utilizing bounding box projection and Fibonacci direction sampling techniques, and introduces a tractable polynomial representation constructed through an implicit polynomial fitting approach. This enables a conservative yet sufficiently accurate analytical approximation of the FOR with a relatively small data set, ensuring system feasibility without excessive conservatism. 

Additionally, we construct an analytical cost function associated with the FOR, enabling the economic valuation and seamless integration of DS flexibility into TSO-level decision-making processes. To operationalize this integration, we develop a FOR-based AC-OPF framework. Within this scheme, the TSO determines the optimal  operating point at the PCC using the analytical models provided by DSs. Subsequently, each DSO performs local FPU dispatch by solving its own AC-OPF based on the TSO’s decision. This single-round coordination mechanism eliminates the need for computationally intensive disaggregation or iterative coordination. 

We benchmark the proposed method against the standard AC-OPF using large-scale, meshed DSs integrated into TSs. Across all test cases, the proposed method achieves average cost deviations below 0.06\% while delivering computational speedups of up to 58\%. The results validate the method’s effectiveness and scalability, demonstrating both high accuracy and computational efficiency. Consequently, DS flexibility can be efficiently integrated into power system operation, while addressing the data privacy concern among stakeholders.

\vspace{-0.3cm}

\bibliographystyle{IEEEtran}
\bibliography{main.bib}

\end{document}